\documentclass[12pt]{article}
\usepackage{amsmath,amssymb}
\usepackage{graphicx,psfrag,epsf}
\usepackage{enumerate}
\usepackage{natbib}
\usepackage{url} 
\usepackage{mathtools}
\usepackage{booktabs}
\usepackage{color}
\usepackage{adjustbox}

\DeclareFontFamily{OT1}{pzc}{}
\DeclareFontShape{OT1}{pzc}{m}{it}{<-> s * [1.10] pzcmi7t}{}
\DeclareMathAlphabet{\mathpzc}{OT1}{pzc}{m}{it}

\newcommand{\bbeta}{{\mbox{\boldmath $\beta$}}}

\newcommand{\bepsilon}{{\mbox{\boldmath $\epsilon$}}}

\newcommand{\x}{{\bf{x}}}
\newcommand{\Y}{{\bf{Y}}}
\newcommand{\X}{{\bf{X}}}
\newcommand{\V}{{\bf{V}}}

\newcommand{\bmu}{{\mbox{\boldmath $\mu$}}}

\let\code=\texttt
\let\proglang=\textsf
\newcommand{\pkg}[1]{{\normalfont\fontseries{b}\selectfont #1}}

\newcommand{\SSE}{\mathsf{SSE}}
\newcommand{\AIC}{\mathsf{AIC}}
\newcommand{\BIC}{\mathsf{BIC}}
\newcommand{\DIC}{\mathsf{DIC}}

\newcommand{\diff}{\mathrm{d}\!}

\begin{document}

\def\spacingset#1{\renewcommand{\baselinestretch}%
{#1}\small\normalsize} \spacingset{1}

\title{Statistical Methods and Computing for Big Data}

\author{Chun Wang, Ming-Hui Chen, Elizabeth Schifano, Jing Wu, and Jun Yan}
\date{\today}

\maketitle

\begin{abstract}
Big data are data on a massive scale in terms of volume, intensity,
and complexity that exceed the capacity of standard analytic tools.
They present opportunities as well as challenges to statisticians.
The role of computational statisticians in scientific discovery from 
big data analyses has been under-recognized even by peer statisticians.
This article summarizes recent methodological and software
developments in statistics that address the big data challenges.
Methodologies are grouped into three classes: subsampling-based,
divide and conquer, and online updating for stream data.
As a new contribution, the online updating approach is 
extended to variable selection with commonly used criteria, and their 
performances are assessed in a simulation study with stream data.
Software packages are summarized with focuses on the open source 
\proglang{R} and \proglang{R} packages, covering recent tools that 
help break the barriers of computer memory and computing power. 
Some of the tools are illustrated in a case study with a logistic
regression for the chance of airline delay.
\end{abstract}




\section{Introduction}
\label{sec:intro}
A 2011 McKinsey report predicted shortage of talent necessary for
organizations to take advantage of big data \citep{Many:etal:big:2011}.
Data now stream from daily life thanks to technological advances, and
big data has indeed become a big deal \citep[e.g.,][]{Shaw:why:2014}.
In the President's Corner of the June 2013 issue of
AMStat News, the three presidents (elect, current, and past) of the
American Statistical Association (ASA) wrote an article titled
``The ASA and Big Data'' \citep{Sche:Davi:Rodr:asa:2013}.
This article echos the June 2012 column of \citet{Rodr:big:2012}
on the recent media focus on big data, and discusses on what the
statistics profession needs to do in response to the fact that
statistics and statisticians are missing from big data discussions.
In the followup July 2013 column, president Marie Davidian further
raised the issues of statistics not being recognized as data science
and mainstream academic statisticians being left behind by the
rise of big data \citep{Davi:aren:2013}.
A white paper prepared by a working group of the ASA called for more
ambitious efforts from statisticians to work together with researchers
in other fields on national research priorities in order to achieve
better science more quickly \citep{Rudi:etal:disc:2014}.
The same concern was expressed in a 2014 president's address of
the Institute of Mathematical Statistics (IMS) \citep{Yu:let:2014}.
President Bin Yu of the IMS called for statisticians to own
Data Science by working on real problems such as those from
genomics, neuroscience, astronomy, nanoscience, computational social
science, personalized medicine/healthcare, finance, and government;
relevant methodology/theory will follow naturally.

Big data in the media or the business world may mean differently than
what are familiar to academic statisticians \citep{Jord:Lin:stat:2014}.
Big data are data on a massive scale in terms of volume, intensity,
and complexity that exceed the ability of standard software tools to
manage and analyze \citep[e.g.,][]{Snij:Matz:Reip:big:2012}.
The origin of the term ``big data'' as it is understood today
has been traced back in a recent study \citep{Dieb:pers:2012} to
lunch-table conversations at Silicon Graphics in the mid-1990s,
in which John Mashey figured prominently \citep{Mash:big:1998}.
Big data are generated by countless online interactions among people,
transactions between people and systems, and sensor-enabled machinery.
Internet search engines (e.g., Google and YouTube) and social network
tools (e.g., Facebook and Twitter) generate billions of activity data per day.
Rather than Gigabytes and Terabytes, nowadays, the data produced are estimated
by zettabytes, and are growing 40\% every day \citep{Fan:Bife:mini:2013}.
In the big data analytics world, a 3V definition by \citet{Lane:3-D:2001}
is widely accepted: volume (amount of data), velocity (speed of data in
and out), and variety (range of data types and sources).
High variety brings nontraditional or even unstructured data types,
such as social network sentiments and internet map usage, which calls
for new, creative ways to understand the structure of data and even to
ask intelligent research questions \citep[e.g.,][]{Jord:Lin:stat:2014}.
High volume and high velocity may bring noise accumulation, spurious
correlation and incidental homogeneity, creating issues in computational
feasibility and algorithmic stability \citep{Fan:Han:Liu:chal:2014}.

Notwithstanding that new statistical thinking and methods are needed
for the high variety aspect of big data, our focus is on fitting
standard statistical models to big data whose size exceeds the
capacity of a single computer from its high volume and high velocity.
There are two computational barriers for big data analysis:
1) the data can be too big to hold in a computer's memory; and
2) the computing task can take too long to wait for the results.
These barriers can be approached either with newly developed
statistical methodologies and/or computational methodologies.
Despite the impression that statisticians are left behind in media
discussions or governmental summits on big data, some statisticians
have made important contributions and are pushing the frontier.
Sound statistical procedures that are scalable computationally
to massive datasets have been proposed \citep{Jord:on:2013}.
Examples are subsampling-based approaches 
\citep{Klei:Talw:Sark:Jord:scal:2014, Ma:Maho:Yu:stat:2013,
Lian:etal:resa:2013, Macl:Adams:fire:2014},
divide and conquer approaches \citep{Lin:Xi:aggr:2011,
  Chen:Xie:spli:2014, Song:Lian:spli:2014, Neis:Wang:Xing:asym:2013},
and online updating approaches \citep{Schifano2015}.
From a computational perspective, much effort has been put into
the most active, open source statistical environment, \proglang{R} \citep{R}.
Statistician \proglang{R} developers are relentless in their drive to
extend the reach of \proglang{R} into big data \citep{Rick:stat:2013}.
Recent UseR! conferences had many presentations that directly
addressed big data, including a 2014 keynote lecture by John Chambers,
the inventor of the \proglang{S} language  \citep{Cham:inte:2014}.
Most cutting edge methods are first and easily implemented in \proglang{R}.
Given the open source nature of \proglang{R} and the active 
recent development, our focus on software for big data will 
be on \proglang{R} and \proglang{R} packages.
Revolution \proglang{R} Enterprise (\proglang{RRE}) is a commercialized 
version of \proglang{R}, but it offers free academic use, 
so it is also included in our case study and benchmarked.
Other commercial software such as \proglang{SAS}, \proglang{SPSS}, and
\proglang{MATLAB} will be briefly touched upon for completeness.

The rest of the article is organized as follows.
Recent methodological developments in statistics on 
big data are summarized in Section~\ref{sec:meth}.
Updating formulas for commonly used variable selection criteria 
in the online setting are developed and their performances 
studied in a simulation study in Section~\ref{sec:varsel}.
Resources from open source software \proglang{R} for analyzing
big data with classical models are summarized in Section~\ref{sec:R}.
Commercial software products are presented in Section~\ref{sec:comm}.
A case study on a logistic model for the chance of airline 
delay is presented in Section~\ref{sec:case}.
A discussion concludes in Section~\ref{sec:disc}.

\section{Statistical Methods}
\label{sec:meth}

The recent methodologies for big data can be loosely grouped into three
categories: resampling-based, divide and conquer, and online updating.
To put the different methods in a context, consider a dataset with
$n$ independent and identically distributed observations, where $n$ is
too big for standard statistical routines such as logistic regression.

\subsection{Subsampling-Based Methods}
\label{sect:stat:resamp}

\subsubsection{Bags of Little Bootstrap}
\citet{Klei:Talw:Sark:Jord:scal:2014} proposed the bags of 
little bootstrap (BLB) approach that provides both point estimates
and quality measures such as variance or confidence intervals.
It is a combination of subsampling \citep{Poli:Roma:Wolf:subs:1999},
the $m$-out-of-$n$ bootstrap \citep{Bick:Gotz:van:resa:1997}, and
the bootstrap \citep{Efro:boot:1979} to achieve computational efficiency.
BLB consists of the following steps.
First, draw $s$ subsamples of size $m$ from the original data of size $n$.
For each of the $s$ subsets, draw $r$ bootstrap samples of size
$n$ instead of $m$, and obtain the point estimates and their quality
measures (e.g., confidence interval) from the $r$ bootstrap sample.
Then, the $s$ bootstrap point estimates and quality measures are
combined (e.g., by average) to yield the overall point estimates
and quality measures.
In summary, BLB has two nested procedures:
the inner procedure applies the bootstrap to a subsample, and the
outer procedure combines these multiple bootstrap estimates.
The subsample size $m$ was suggested to be $n^{\gamma}$ with
$\gamma \in [0.5, 1]$ \citep{Klei:Talw:Sark:Jord:scal:2014},
a much smaller number than $n$.
Although the inner bootstrap procedure conceptually generates
multiple resampled data of size $n$, what is really needed in the
storage and computation is a sample of size $m$ with a weight vector.
In contrast to subsampling and the $m$-out-of-$n$ bootstrap, there
is no need for an analytic correction (e.g., $\sqrt{m/n}$) to
rescale the confidence intervals from the final result.
The BLB procedure facilitates distributed computing by letting
each subsample of size $m$  be processed by a separate processor.
\citet{Klei:Talw:Sark:Jord:scal:2014} proved the consistency of BLB
and provided high order correctness.
Their simulation study showed good accuracy, convergence rate
and the remarkable computational efficiency.

\subsubsection{Leveraging}
\citet{Ma:Sun:leve:2015} proposed to use leveraging to facilitate
scientific discoveries from big data using limited computing resources.
In a leveraging method, one samples a small proportion of the data with 
certain weights (subsample) from the full sample, and then performs intended 
computations for the full sample using the small subsample as a surrogate. 
The key to success of the leveraging methods is to construct the weights,
the nonuniform sampling probabilities, so that influential data points 
are sampled with high probabilities \citep{Ma:Maho:Yu:stat:2013}.
Leveraging methods are different from the traditional subsampling or
$m$-out-of-$n$ bootstrap in that 
1) they are used to achieve feasible 
computation even if the simple analytic results are available;
2) they enable visualization of the data when visualization of the
full sample is impossible; and 
3) they usually use unequal sampling probabilities for subsampling data.
This approach is quite unique in allowing pervasive access to extract
information from big data without resorting to high performance computing.

\subsubsection{Mean Log-likelihood}
\citet{Lian:etal:resa:2013} proposed a resampling-based stochastic
approximation approach with an application to big geostatistical data.
The method uses Monte Carlo averages calculated from subsamples
to approximate the quantities needed for the full data.
Motivated from minimizing the Kullback--Leibler (KL) divergence, they
approximate the KL divergence by averages calculated from subsamples.
This leads to a maximum mean log-likelihood estimation method.
The solution to the mean score equation is obtained from a stochastic
approximation procedure, where at each iteration, the current estimate
is updated based on a subsample of size $m$ drawn from the full data.
As $m$ is much smaller than $n$, the method is scalable to big data.
\citet{Lian:etal:resa:2013} established the consistency and asymptotic
normality of the resulting estimator under mild conditions.
In a simulation study, the convergence rate of the method was
almost independent of $n$, the sample size of the full data.

\subsubsection{Subsampling-Based MCMC}
As a popular general purpose tool for Bayesian inference, 
Markov chain Monte Carlo (MCMC)
for big data is challenging because of the prohibitive cost of 
likelihood evaluation of every datum at every iteration.
\citet{Lian:Kim:boot:2013} extended the mean log-likelihood 
method to a bootstrap Metropolis--Hastings (MH) algorithm in MCMC.
The likelihood ratio of the proposal and current estimate in the
MH ratio is replaced with an approximation from the mean 
log-likelihood based on $k$ bootstrap samples of size $m$.
The algorithm can be implemented exploiting the embarrassingly parallel
structure and avoids repeated scans of the full dataset in iterations.
\citet{Macl:Adams:fire:2014} proposed an auxiliary variable MCMC
algorithm called Firefly Monte Carlo (FlyMC) that only queries 
the likelihoods of a potentially small subset of the data at each
iteration yet simulates from the exact posterior distribution.
For each data point, a binary auxiliary variable and a strictly
positive lower bound of the likelihood contribution are introduced.
The binary variable for each datum effectively turn on and off 
data points in the posterior, hence the ``firefly" name. 
The probability of turning on each datum depends on the ratio 
of its likelihood contribution and the introduced lower bound.
The computational gain depends on that the lower bound is tight
enough and that simulation of the auxiliary variables is cheap enough.
Because of the need to hold the whole data in computer memory,
the size of the data this method can handle is limited.

The pseudo-marginal Metropolis--Hasting algorithm replaces the
intractable target (posterior) density in the MH algorithm with an
unbiased estimator \citep{Andr:Robe:pseu:2009}.
The the log-likelihood is estimated by an unbiased subsampled 
version, and an unbiased estimator of the likelihood is obtained 
by correcting the bias of the exponentiation of this estimator.
\citet{Quir:Vill:Kohn:spee:2014} proposed subsampling the data 
using probability proportional-to-size (PPS) sampling to obtain
an approximately unbiased estimate of the likelihood which is 
used in the M-H acceptance step.
The subsampling approach was further improved in 
\citet{Quir:Vill:Kohn:scal:2015} using the efficient and robust 
difference estimator form the survey sampling literature.

\subsection{Divide and Conquer}
\label{sect:stat:dc}

A divide and conquer algorithm (which may appear under other names
such as divide and recombine, split and conquer, or split and merge)
generally has three steps:
1) partitions a big dataset into $K$ blocks;
2) processes each block separately (possibly in parallel);
and 3) aggregates the solutions from each block
to form a final solution to the full data.

\subsubsection{Aggregated Estimating Equations}
For a linear regression model, the least squares estimator for the
regression coefficient $\beta$ for the full data can be expressed
as a weighted average of the least squares estimator for each block
with weight being the inverse of the estimated variance matrix.
The success of this method for linear regression depends on the linearity
of the estimating equations in $\beta$ and that the estimating equation
for the full data is a simple summation of that for all the blocks.
For general nonlinear estimating equations, \citet{Lin:Xi:aggr:2011}
proposed a linear approximation of the estimating equations with
the Taylor expansion at the solution in each block, and, hence,
reduce the nonlinear estimating equation to the linear case so that
the solutions to all the blocks are combined by a weighted average.
The weight of each block is the slope matrix of the estimating function
at the solution in that block, which is the Fisher information or
inverse of the variance matrix if the equations are score equations.
\citet{Lin:Xi:aggr:2011} showed that, under certain technical
conditions including $K = O(n^{\gamma})$ for some $\gamma \in (0, 1)$,
the aggregated estimator has the same limit as the estimator from
the full data.

\subsubsection{Majority Voting}
\citet{Chen:Xie:spli:2014} consider a divide and conquer approach for
generalized linear models (GLM) where both the sample size $n$ and 
the number of covariates $p$ are large, by incorporating variable
selection via penalized regression into a subset processing step.
More specifically, for $p$ bounded or increasing to infinity slowly, 
($p_n$ not faster than $o(e^{n_k})$, while model size may increase at 
a rate of $o(n_k)$), they propose to first randomly split the data
of size $n$ into $K$ blocks (size $n_k=O(n/K)$). 
In step~2, penalized regression is applied to each block separately with a 
sparsity-inducing penalty function satisfying certain regularity conditions. 
This approach can lead to differential variable selection among the 
blocks, as different blocks of data may result in penalized estimates 
with different non-zero regression coefficients. 
Thus, in step~3, the results from the $K$ blocks are combined by 
majority vote to create a combined estimator. 
The implicit assumption is that real effects should be found 
persistently and therefore should be present even under perturbation 
by subsampling \citep[e.g.][]{meinsh.buhl:2010}.
The derivation of the combined estimator in step~3 stems from 
ideas for combining confidence distributions in meta-analysis 
\citep{singh.xie.straw:2005, xie.singh.straw:2011},
where one can think of the $K$ blocks as $K$ independent and
separate analyses to be combined in a meta-analysis.
The authors show under certain regularity conditions that their 
combined estimator in step~3 is model selection consistent, 
asymptotically equivalent to the penalized estimator that would
result from using all of the data simultaneously, and achieves 
the oracle property when it is attainable for the penalized 
estimator from each block \citep[see e.g.,][]{fan.lv:2011}.  
They additionally establish an upper bound for the expected number
of incorrectly selected variables and a lower bound for the expected
number of correctly selected variables.

\subsubsection{Screening with Ultrahigh Dimension}
Instead of dividing the data into blocks of observations in step~1,
\citet{Song:Lian:spli:2014} proposed a split-and-merge (SAM) method 
that divides the data into subsets of covariates for variable selection
in ultrahigh dimensional regression from the Bayesian perspective.  
This method is particularly suited for big data where the number
of covariates $P_n$ is much larger than the sample size $n$, 
$P_n \gg n$, and possibly increasing with $n$.
In step~2, Bayesian variable selection is separately performed on
each lower dimensional subset, which facilitates parallel processing.
In step~3, the selected variables from each subset are aggregated,
and Bayesian variable selection is applied on the aggregated data.
The embarrassingly parallel structure in step~2 makes the SAM method
applicable to big data problems with millions or more predictors.  
Posterior consistency is established for correctly specified 
models and for misspecified models, the latter of which is necessary
because it is quite likely that some true predictors are missing.
With correct model specification, true covariates will be identified 
as the sample size becomes large; under misspecified models, all
predictors correlated with the response variable will be identified.
Compared with the sure independence screening (SIS) approach
\citep{Fan:Lv:sure:2008}, the method uses the joint information
of multiple predictors in predictor screening while SIS only 
uses the marginal information of each predictor.  
Their numerical results show that the SAM approach outperforms
competing methods for ultrahigh dimensional regression.

\subsubsection{Parallel MCMC}
In the Bayesian framework, it is natural to partition the data 
into $K$ subsets and run parallel MCMC on each one of them.
The prior distribution for each subset is often obtained by
taking a power $1/K$ of the prior distribution for whole data
in order to preserve the total amount of prior information
(which may change the impropriety of the prior).
MCMC is run independently on each subset with no 
communications between subsets (and, thus, embarrassingly 
parallel), and the resulting samples are combined to approximate 
samples from the full data posterior distribution.
\citet{Neis:Wang:Xing:asym:2013} proposed to use kernel density
estimators of the posterior density for each data subset, and estimate
the full data posterior by multiplying the subset posterior densities
together. This method is asymptotically exact in the sense of 
being converging in the number of MCMC iterations. 
\citet{Wang:etal:para:2015} replaced the kernel estimator of
\citet{Neis:Wang:Xing:asym:2013} with a random partition tree 
histogram, which uses the same block partition across all terms in 
the product representation of the posterior to control the number of
terms in the approximation such that it does not explode with $m$. 
\citet{Scot:etal:baye:2013} proposed a consensus Monte Carlo
algorithm, which produces the approximated full data posterior
using weighted averages over the subset MCMC samples.
The weight used (for Gaussian models) for each subset is the 
inverse of the variance-covariance matrix of the MCMC samples.
The method is effective when the posterior is close to Gaussian but
may cause bias when the distribution is skewed or has multi-modes. 
The consensus Monte Carlo principal is approached from a 
variational perspective by \citet{Robi:Ange:Jord:vari:2015}.
The embarrassingly parallel feature of these methods facilitates their 
implementation in the MapReduce framework that exploits the 
division and recombination strategy \citep{Dean:Ghem:mapr:2008}. 
The final recombination step is implemented in \proglang{R} package 
\texttt{parallelMCMCcombine} \citep{Miro:Conl:para:2014}.

Going beyond embarrassingly parallel MCMC remains challenging
because of storage issues and communication overheads.
General strategies for parallel MCMC such as multiple-proposal 
MH algorithm \citep{Cald:gene:2014} and population MCMC
\citep{Song:Wu:Lian:weak:2014} mostly require full data at each node.

\subsection{Online Updating for Stream Data}
In some applications, data come in streams or large chunks, and a
sequentially updated analysis is desirable without storing the data.
Motivated from a Bayesian inference perspective, \citet{Schifano2015}
extends the work of \citet{Lin:Xi:aggr:2011} in a few important ways.
First, they introduce divide-and-conquer-type variance estimates of
regression parameters in the linear model and estimating equation settings.
These estimates of variability allow for users to make inferences about
the true regression parameters based upon the previously developed
divide-and-conquer point estimates of the regression parameters.
Second, they develop iterative estimating algorithms and statistical
inferences for linear models and estimating equations that update as
new data arrive. Thus, while the divide-and-conquer setting is quite
amenable to parallel processing for each subset, the online-updating
approach for data streams is inherently sequential in nature.
Their algorithms were designed to be computationally efficient and minimally
storage-intensive, as they assume no access/storage of the historical data.
Third, the authors address the issue of possible rank deficiencies
when dealing with blocks of data, and the uniqueness properties of the
combined and cumulative estimators when using a generalized inverse.
The authors also provide methods for assessing goodness of fit in the
linear model setting, as standard residual-based diagnostics cannot
be performed with the cumulative data without access to historical data.
Instead, they propose outlier tests relying on predictive residuals, which
are based on the predictive values computed from the cumulative estimate
of the regression coefficients attained at the previous accumulation point.
Additionally, they introduce a new online-updated estimator of the
regression coefficients and corresponding estimator of the standard error
in the estimating equation setting that takes advantage of information
from the previous data.  They show theoretically that this new estimator,
the cumulative updated estimating equation (CUEE) estimator, is
asymptotically consistent, and show empirically that the CUEE estimator
is less biased in their finite sample simulations than the cumulatively
estimated version of the estimator of \citet{Lin:Xi:aggr:2011}.

\section{Criterion-Based Variable Selection with Online Updating}
\label{sec:varsel}

To the best of our knowledge, criterion-based variable selection
has not yet been considered in the online updating context. 
This problem is well worth investigating especially
when access/storage of the historical data is limited. 
Suppose that we have $K$ blocks of data in a sequence with $\Y_k$, 
$\X_k$, and $n_k$ being the $n_k$-dimensional vector of responses,
the $n_k \times (p + 1)$ matrix of covariates, and the sample size, 
respectively, for the $k_{th}$ block, $k = 1, \ldots, K$,
such that $\Y = (Y'_1, Y'_2, \dots, Y'_K )'$ and 
$\X = (\X'_1, \dots, \X'_k)'$.
Consider the standard linear regression model for the whole data
with sample size $n = \sum_{i=1}^k n_k$,
\[
\Y = \X \bbeta + \bepsilon,
\]
where $\bbeta$ is the regression coefficient vector, 
and $\bepsilon$ is a normal random vector with 
mean zero and variance $\theta I_n$.
Let $\mathcal{M}$ denote the model space.
We enumerate the models in $\mathcal{M}$  by $m=1, 2, ..., 2^p$,
where $2^p$ is the dimension of $\mathcal{M}$.
For the full model, the least squares estimate of $\bbeta$ and
the sum of squared errors based on the $k$th subset is given by
$\hat{\bbeta}_{n_k,k}=(\X'_k\X_k)^{-}\X'_k\Y_k$ and $\SSE_{n_k, k}$.
In the sequential setting, we only need to store and update the 
cumulative estimates at each $k$ \citep[see, e.g.][]{Schifano2015}.

Let $\bbeta^{(m)}_{k}=(\beta^{(m)}_{0}, \beta^{(m)}_{1},  \ldots, \beta^{(m)}_{{p_m}})'$ 
and $\SSE^{(m)}_{k}$ denote the cumulative estimates based on all 
data through subset $k$ for model $m$, 
where $p_m$ is the number of covariates for model $m$.
We further introduce the $(p + 1) \times (p_m +1)$ selection
matrix $P^{(m)} = (e_{m_0}, \quad e_{m_1}, \quad \dots \quad e_{m_{p_m}})$,
where $e_{m_0}$ is a vector with length $(p+1)$ and the first element as 1, 
and $e_{m_j}$ denotes a vector of length $(p+1)$ with 1 
in the $m_{j}$th position and 0 in every other position for all $j > 0$. 
Here $(m_1, ..., m_{p_m})$ are not necessarily in sequence, but represents the index
of selected variables in the full design matrix $X_k$. 
Define $\X^{(m)}_{k} = \X_kP^{(m)}$. Update a $(p_m + 1) \times (p_m + 1)$ matrix
\begin{equation*} 
V^{(m)}_{k}= \X^{(m)'}_{k} \X^{(m)}_{k} + V^{(m)}_{k-1}, 
\end{equation*}  
where $V^{(m)}_0=0$, and a $(p_m + 1) \times 1$ vector
\begin{equation*} 
A^{(m)}_k = \X^{(m)'}_{k}\X_k\hat{\bbeta}_{n_k,k}+V^{(m)}_{k-1}\hat{\bbeta}^{(m)}_{k-1},
\end{equation*} 
where $\hat{\bbeta}^{(m)}_{0} = 0$.
After some algebra, we have
\begin{equation*}
\hat{\bbeta}^{(m)}_{k} = (V^{(m)}_k)^{-1} A^{(m)}_k,
\end{equation*}
and 
\begin{align*}
\SSE^{(m)}_{k} &= \SSE_{n_k k}+\hat{\bbeta}_{n_k k}'\X'_k\X_k\hat{\bbeta}_{n_k k}+\hat{\bbeta}^{(m)'}_{k-1}V^{(m)}_{k-1}\hat{\bbeta}^{(m)}_{k-1} \\
& \quad - \hat{\bbeta}^{(m)'}_{k}V^{(m)}_{k}\hat{\bbeta}^{(m)}_{k} + \SSE^{(m)}_{k-1}.
\end{align*}

With $\sigma$ unknown, letting
\begin{align*}
B_k^{(m)} & = n \log \frac{2\pi \SSE^{(m)}_{k}}{n - p_m - 1},
\end{align*}
the Akaike information criterion (AIC) and 
Bayesian information criterion (BIC) are updated by
\begin{align*}
\AIC_k^{(m)} &=  B_k^{(m)}  + n + p_m + 1,\\
\BIC_k^{(m)} &=  B_k^{(m)}  + n - p_m - 1 + (p_{m}+1)\log n.
\end{align*}

To study the Bayesian variable selection criteria, assume a joint 
conjugate prior for $(\bbeta^{(m)},{\theta}^{(m)})$ as follows:
$\bbeta^{(m)} | {\theta}^{(m)}$ follows normal distribution with mean $\bmu_o$, 
and precision matrix $\V_0$, ${\theta}^{(m)}$ follows 
Inverse Gamma distribution with shape parameter $\nu_0/2$ and 
scale parameter $\tau_0/2$, e.g,
\begin{align*}
  \pi( \bbeta^{(m)},{\theta}^{(m)}| & \bmu_0, \V_0, \nu_0, \tau_0) \\
  & = \pi(\bbeta^{(m)}|{\theta}^{(m)},\bmu_0, \V_0) \pi({\theta}^{(m)}|\nu_0,\tau_0),
\end{align*}
where $\bmu_0$ is a prespecified $(p_m+1)$-dimensional vector,
$\V_0$ is a $(p_m+1)\times (p_m+1)$ positive definite matrix,
$\nu_0 > 0$, $\tau_0 > 0$.
It can be shown that the deviance information criterion ($\DIC$)
\citep{Spie:etal:baye:2002} is updated by 
\begin{equation*}
\DIC_k^{(m)} = n \log\frac{\pi(n-2) \SSE_k^{(m)}}{2} + 2n\psi(\frac{n}{2}) + 2p_m +n + 4,
\end{equation*}
where $\psi(x)=\diff \log \Gamma(x) / \diff x$ is the digamma function.

We examined the performance of AIC, BIC and DIC under
the online updating scenario in a simulation study.
Each dataset was generated from linear model
$y_i= \x_i'\bbeta + \epsilon_i,$  where 
$\epsilon_i$'s were independently generated from $N(0, 100)$,
$x_i = (1, x_{i1}, x_{i2}, x_{i3}, x_{i4})$ were identically distributed
random vectors from a multivariate normal distribution with mean
$(1, 0, 0, 0, 0)$ and marginal variances $(0, 16, 9, 0.3, 3)$. 
Two correlation structures of $(x_{i1}, x_{i2}, x_{i3}, x_{i4})$
were considered: 1) independent and 
2) AR(1) with correlation coefficient 0.9.
Four different models as determined by the nonzeroness of $\bbeta$
were considered: $(-1, 3, 0, 0, 0)$, $(-1, 3, 0, -1.5, 0)$,
$(-1, 3, 2, -1.5, 0)$, and $(-1, 3, 2, -1.5, 1)$.
The corresponding signal-to-noise ratios were
1.44, 1.45, 1.81, and 1.83 in the independent case and 
1.44, 1.29, 2.85, and 3.33 under the dependent case.
The sample size of each block was set as $n_k=100$.
The performance of the criteria was investigated with the 
cumulative estimates at block $k \in \{2, 25, 100\}$.
For each scenario, 10,000 independent datasets were generated.

\begin{table*}[tbp]
  \caption{Percentages of the simulations that identify the variables indicated on the left for various number of blocks ($k$), subset sample sizes ($n_k=100$) and correlation within the design matrix $\X$ (independent or dependent). }
  \label{tab:sim}
  \centering
\begin{adjustbox}{max width=0.85\textwidth}
\begin{tabular}{l ccc ccc ccc ccc ccc ccc}
\toprule
\cmidrule(lr){2-10} \cmidrule(lr){11-19}
& \multicolumn{9}{c}{independent} & \multicolumn{9}{c}{dependent} \\
\cmidrule(lr){2-10} \cmidrule(lr){11-19}
True   & \multicolumn{3}{c}{$k=2$} &  \multicolumn{3}{c}{$k=25$}  & \multicolumn{3}{c}{$k=100$} &  \multicolumn{3}{c}{$k=2$} &   \multicolumn{3}{c}{$k=25$}  & \multicolumn{3}{c}{$k=100$}\\
\cmidrule(lr){2-4}\cmidrule(lr){5-7}\cmidrule(lr){8-10}
\cmidrule(lr){11-13}\cmidrule(lr){14-16}\cmidrule(lr){17-19}
Model & AIC &BIC &DIC& AIC &BIC &DIC& AIC &BIC &DIC &AIC &BIC &DIC& AIC &BIC &DIC& AIC &BIC &DIC\\
\midrule
\multicolumn{19}{l}{$\bbeta=(-1, 3, 0, 0, 0)$, signal-to-noise ratios are 1.44 for both independent and dependent.}\\
none    &0	&0	&0	&0	&0	&0	&0	&0	&0	&0	&0	&0	&0	&0	&0	&0	&0	&0\\
$\bf(x_1)$   &\bf59	&\bf93	&\bf59	&\bf60	&\bf98	&\bf60	&\bf59	&\bf99	&\bf59	&\bf63	&\bf94	&\bf62	&\bf64	&\bf99	&\bf64	&\bf64	&\bf99	&\bf64\\
($x_2$)     &0	&0	&0	&0	&0	&0	&0	&0	&0	&0	&0	&0	&0	&0	&0	&0	&0	&0\\
($x_1, x_2$)  &11	&2	&11	&11	&1	&11	&12	&0	&12	&10	&2	&10	&9	&1	&9	&10	&0	&10\\
($x_3$)      &0	&0	&0	&0	&0	&0	&0	&0	&0	&0	&0	&0	&0	&0	&0	&0	&0	&0\\
($x_1, x_3$)  &11	&2	&11	&11	&1	&11	&11	&0	&11	&8	&2	&8	&8	&0	&8	&8	&0	&8\\
($x_2, x_3$)    &0	&0	&0	&0	&0	&0	&0	&0	&0	&0	&0	&0	&0	&0	&0	&0	&0	&0\\
($x_1, x_2, x_3$) &2	&0	&3	&2	&0	&2	&2	&0	&2	&4	&0	&4	&3	&0	&3	&3	&0	&3\\
($x_4$)      &0	&0	&0	&0	&0	&0	&0	&0	&0	&0	&0	&0	&0	&0	&0	&0	&0	&0\\
($x_1, x_4$)  &11	&2	&11	&11	&0	&11	&11	&0	&11	&9	&2	&9	&8	&0	&9	&8	&0	&8\\
($x_2, x_4$)   &0	&0	&0	&0	&0	&0	&0	&0	&0	&0	&0	&0	&0	&0	&0	&0	&0	&0\\
($x_1, x_2, x_4$)  &2	&0	&2	&2	&0	&2	&2	&0	&2	&3	&0	&3	&3	&0	&3	&3	&0	&3\\
($x_3, x_4$)    &0	&0	&0	&0	&0	&0	&0	&0	&0	&0	&0	&0	&0	&0	&0	&0	&0	&0\\
($x_1, x_3, x_4$) &2	&0	&2	&2	&0	&2	&2	&0	&2	&4	&0	&4	&4	&0	&4	&4	&0	&4\\
($x_2, x_3, x_4$)  &0	&0	&0	&0	&0	&0	&0	&0	&0	&0	&0	&0	&0	&0	&0	&0	&0	&0\\
($x_1, x_2, x_3, x_4$)&1	&0	&1	&0	&0	&0	&0	&0	&0	&1	&0	&1	&1	&0	&1	&1	&0	&1\\
[5pt]
\multicolumn{19}{l}{$\bbeta=(-1, 3, 0, -1.5, 0)$, signal-to-noise ratios are 1.45 for independent and 1.29 for dependent.}\\
none        &0	&0	&0	&0	&0	&0	&0	&0	&0	&0	&0	&0	&0	&0	&0	&0	&0	&0\\
($x_1$)     &42	&83	&42	&0	&9	&0	&0	&0	&0	&55	&89	&55	&10	&60	&10	&0	&3	&0\\
($x_2$)     &0	&0	&0	&0	&0	&0	&0	&0	&0	&0	&0	&0	&0	&0	&0	&0	&0	&0\\
($x_1, x_2$)    &8	&2	&8	&0	&0	&0	&0	&0	&0	&11	&3	&11	&10	&4	&10	&1	&2	&1\\
($x_3$)      &0	&0	&0	&0	&0	&0	&0	&0	&0	&0	&0	&0	&0	&0	&0	&0	&0	&0\\
$\bf(x_1, x_3)$   &\bf28	&\bf12	&\bf27	&\bf71	&\bf90	&\bf71	&\bf70	&\bf100	&\bf70	&\bf13	&\bf4	&\bf13	&\bf50	&\bf30	&\bf50	&\bf69	&\bf90	&\bf69\\
($x_2, x_3$)    &0	&0	&0	&0	&0	&0	&0	&0	&0	&0	&0	&0	&0	&0	&0	&0	&0	&0\\
($x_1, x_2, x_3$) &6	&0	&6	&13	&0	&13	&14	&0	&14	&4	&0	&4	&6	&0	&6	&12	&0	&12\\
($x_4$)      &0	&0	&0	&0	&0	&0	&0	&0	&0	&0	&0	&0	&0	&0	&0	&0	&0	&0\\
($x_1, x_4$)   &8	&2	&8	&0	&0	&0	&0	&0	&0	&10	&3	&10	&14	&6	&14	&3	&5	&3\\
($x_2, x_4$)    &0	&0	&0	&0	&0	&0	&0	&0	&0	&0	&0	&0	&0	&0	&0	&0	&0	&0\\
($x_1, x_2, x_4$) &2	&0	&2	&0	&0	&0	&0	&0	&0	&3	&0	&3	&2	&0	&2	&2	&0	&2\\
($x_3, x_4$)    &0	&0	&0	&0	&0	&0	&0	&0	&0	&0	&0	&0	&0	&0	&0	&0	&0	&0\\
($x_1, x_3, x_4$)  &6	&0	&6	&13	&0	&13	&13	&0	&13	&4	&0	&5	&6	&0	&6	&11	&0	&11\\
($x_2, x_3, x_4$)  &0	&0	&0	&0	&0	&0	&0	&0	&0	&0	&0	&0	&0	&0	&0	&0	&0	&0\\
($x_1, x_2, x_3, x_4$)&1	&0	&1	&2	&0	&3	&3	&0	&3	&1	&0	&1	&1	&0	&1	&2	&0	&2\\
[5pt]
\multicolumn{19}{l}{$\bbeta=(-1, 3, 2, -1.5, 0)$, signal-to-noise ratios are 1.81 for independent and 2.85 for dependent.}\\
none        &0	&0	&0	&0	&0	&0	&0	&0	&0	&0	&0	&0	&0	&0	&0	&0	&0	&0\\
($x_1$)     &0	&0	&0	&0	&0	&0	&0	&0	&0	&2	&17	&2	&0	&0	&0	&0	&0	&0\\
($x_2$)      &0	&0	&0	&0	&0	&0	&0	&0	&0	&0	&0	&0	&0	&0	&0	&0	&0	&0\\
($x_1, x_2$)  &50	&85	&50	&0	&9	&0	&0	&0	&0	&64	&74	&64	&28	&83	&28	&1	&29	&1\\
($x_3$)     &0	&0	&0	&0	&0	&0	&0	&0	&0	&0	&0	&0	&0	&0	&0	&0	&0	&0\\
($x_1, x_3$)   &0	&0	&0	&0	&0	&0	&0	&0	&0	&3	&2	&3	&0	&0	&0	&0	&0	&0\\
($x_2, x_3$)   &0	&0	&0	&0	&0	&0	&0	&0	&0	&0	&0	&0	&0	&0	&0	&0	&0	&0\\
$\bf(x_1, x_2, x_3)$ &\bf33	&\bf13	&\bf33	&\bf84	&\bf90	&\bf84	&\bf84	&\bf100	&\bf84	&\bf14	&\bf3	&\bf14	&\bf50	&\bf14	&\bf50	&\bf81	&\bf67	&\bf81\\
($x_4$)     &0	&0	&0	&0	&0	&0	&0	&0	&0	&0	&0	&0	&0	&0	&0	&0	&0	&0\\
($x_1, x_4$)   &0	&0	&0	&0	&0	&0	&0	&0	&0	&1	&1	&1	&0	&0	&0	&0	&0	&0\\
($x_2, x_4$)    &0	&0	&0	&0	&0	&0	&0	&0	&0	&0	&0	&0	&0	&0	&0	&0	&0	&0\\
($x_1, x_2, x_4$) &10	&2	&10	&0	&0	&0	&0	&0	&0	&11	&2	&11	&15	&3	&15	&6	&4	&6\\
($x_3, x_4$)   &0	&0	&0	&0	&0	&0	&0	&0	&0	&0	&0	&0	&0	&0	&0	&0	&0	&0\\
($x_1, x_3, x_4$) &0	&0	&0	&0	&0	&0	&0	&0	&0	&1	&0	&1	&0	&0	&0	&0	&0	&0\\
($x_2, x_3, x_4$) &0	&0	&0	&0	&0	&0	&0	&0	&0	&0	&0	&0	&0	&0	&0	&0	&0	&0\\
$(x_1, x_2, x_3, x_4)$ &7	&0	&7	&15	&0	&15	&16	&0	&16	&4	&0	&5	&7	&0	&7	&13	&0	&13\\
[5pt]
\multicolumn{19}{l}{$\bbeta=(-1, 3, 2, -1.5, 1)$, signal-to-noise ratios are 1.84 for independent and 3.33 for dependent.} \\
none        &0	&0	&0	&0	&0	&0	&0	&0	&0	&0	&0	&0	&0	&0	&0	&0	&0  & 0 \\
($x_1$)      &0	&0	&0	&0	&0	&0	&0	&0	&0	&0	&3	&0	&0	&0	&0	&0	&0	&0\\
($x_2$)      &0	&0	&0	&0	&0	&0	&0	&0	&0	&0	&0	&0	&0	&0	&0	&0	&0	&0\\
($x_1, x_2$)  &9	&40	&9	&0	&0	&0	&0	&0	&0	&51	&75	&51	&0	&13	&0	&0	&0	&0\\
($x_3$)      &0	&0	&0	&0	&0	&0	&0	&0	&0	&0	&0	&0	&0	&0	&0	&0	&0	&0\\
($x_1, x_3$)    &0	&0	&0	&0	&0	&0	&0	&0	&0	&4	&6	&4	&0	&0	&0	&0	&0	&0\\
($x_2, x_3$)    &0	&0	&0	&0	&0	&0	&0	&0	&0	&0	&0	&0	&0	&0	&0	&0	&0	&0\\
($x_1, x_2, x_3$)  &6	&6	&6	&0	&0	&0	&0	&0	&0	&7	&1	&7	&0	&0	&0	&0	&0	&0\\
($x_4$)      &0	&0	&0	&0	&0	&0	&0	&0	&0	&0	&0	&0	&0	&0	&0	&0	&0	&0\\
($x_1, x_4$)    &0	&0	&0	&0	&0	&0	&0	&0	&0	&4	&10	&4	&0	&0	&0	&0	&0	&0\\
($x_2, x_4$)    &0	&0	&0	&0	&0	&0	&0	&0	&0	&0	&0	&0	&0	&0	&0	&0	&0	&0\\
($x_1, x_2, x_4$) &50	&47	&50	&0	&9	&0	&0	&0	&0	&24	&4	&25	&51	&80	&51	&11	&65	&11\\
($x_3, x_4$)    &0	&0	&0	&0	&0	&0	&0	&0	&0	&0	&0	&0	&0	&0	&0	&0	&0	&0\\
($x_1, x_3, x_4$)  &0	&0	&0	&0	&0	&0	&0	&0	&0	&0	&0	&0	&0	&0	&0	&0	&0	&0\\
($x_2, x_3, x_4$)  &0	&0	&0	&0	&0	&0	&0	&0	&0	&0	&0	&0	&0	&0	&0	&0	&0	&0\\
$\bf(x_1, x_2, x_3, x_4)$ &\bf34	&\bf7	&\bf34	&\bf100	&\bf91	&\bf100	&\bf100	&\bf100	&\bf100	&\bf10	&\bf1	&\bf10	&\bf48	&\bf7	&\bf48	&\bf89	&\bf35	&\bf89\\
\bottomrule
\end{tabular}
\end{adjustbox}
\end{table*}

The percentages of models selected among the $2^4$ models by 
each of the three criteria are summarized in Table~\ref{tab:sim}.
The entire row in bold represents the true model.
Based on the simulation results, BIC performs extremely well 
when the number of blocks ($k$) is large, which is consistent
with known results that the probability of selecting the
true model by BIC approaches 1 as $n \rightarrow \infty$ 
\citep[e.g.,][]{Schw:esti:1978, Nish:asym:1984}.
The BIC also performs better than AIC and DIC when the
covariates are independent, even for small sample sizes. 
When covariates are highly dependent, AIC and DIC provide
more reliable results when sample size is small.
The performance of AIC and DIC is always very similar. 
The simulation results also confirm the existing theorem that
AIC is not consistent \citep[e.g.,][]{Wood:on:1982}. 
In the big data setting with large sample size, BIC is generally
preferable, especially when the covariates are not highly correlated.

\section{Open Source R and R Packages}
\label{sec:R}

Handling big data is one of the topics of high performance computing.
As the most popular open source statistical software, \proglang{R} and its 
adds-on packages provide a wide range of high performance computing; 
see Comprehensive \proglang{R} Archive Network (CRAN) task view on 
``High-Performance and Parallel Computing with R'' \citep{Edde:cran:2014}.
The focus of this section is on how to break the computer memory barrier 
and the computing power barrier in the context of big data.

\subsection{Breaking the Memory Barrier}
\label{sec:R:mem}

The size of big data is relative to the available computing resources.
The theoretical limit of random access memory (RAM) is determined by 
the width of memory addresses: 4 gigabyte (GB) ($2^{32}$ bytes) for a 32-bit 
computer and 16.8 million terabyte ($2^{64}$ bytes) for a 64-bit computer.
In practice, however, the latter is limited by the physical space
of a computer case, the operating system, and specific software.
Individual objects in \proglang{R} have limits in size too; an \proglang{R} 
user can hardly work with any object of size close to that limit.
\citet{Emer:Kane:dont:2012} suggested that a data set would be 
considered \emph{large} if it exceeds 20\% of RAM on a given machine 
and \emph{massive} if it exceeds 50\%, in which case, even the simplest 
calculation would consume all the remaining RAM.

Memory boundary can be broken with an external memory algorithms (EMA)
\citep[e.g.,][]{Vitt:exte:2001}, which conceptually works by storing
the data on a disk storage (which has a much greater limit than RAM), and 
processing one chunk of it at a time in RAM \citep[e.g.,][]{Rpkg:biglm}.
The results from each chunk will be saved or updated and the process
continues until the entire dataset is exhausted; then, if needed as in an 
iterative algorithm, the process is reset from the beginning of the data.
To implement an EMA for each statistical function, one need to address
1) data management and 2) numerical calculation.

\subsubsection{Data Management}
Earlier solutions to oversize data resorted to relational databases.
This method depends on an external database management system (DBMS)
such as \proglang{MySQL}, \proglang{PostgreSQL}, \proglang{SQLite}, 
\proglang{H2}, \proglang{ODBC}, \proglang{Oracle}, and others.
Interfaces to \proglang{R} are provided through many \proglang{R}
packages such as
\pkg{sqldf} \citep{Rpkg:sqldf}, 
\pkg{DBI} \citep{Rpkg:dbi},
\pkg{RSQLite} \citep{Rpkg:rsqlite}, 
and others.
The database approach requires a DBMS to be installed and maintained,
and knowledge of structured query language (\proglang{SQL}); an exception 
for simpler applications is package \pkg{filehash} \citep{Peng:inte:2006}, 
which comes with a simple key-value database implementation itself.
The numerical functionality of SQL is quite limited, and calculations
for most statistical analyses require copying subsets of the data 
into objects in \proglang{R} facilitated by the interfaces.
Extracting chunks from an external DBMS is computationally much 
less efficient than the more recent approaches discussed below
\citep{Kane:Emer:West:scal:2013}.

Two \proglang{R} packages, \pkg{bigmemory} \citep{Kane:Emer:West:scal:2013}
and \pkg{ff} \citep{Rpkg:ff} provide data structures for massive
data while retaining a look and feel of \proglang{R} objects.
Package \pkg{bigmemory} defines a data structure \code{big.matrix}
for numeric matrices which uses memory-mapped files to allow matrices 
to exceed the RAM size on computers with 64-bit operating systems.
The underling technology is memory mapping on modern operating 
systems that associates a segment of virtual memory in a one-to-one
correspondence with contents of a file.
These files are accessed at a much faster speed than in the database
approaches because operations are handled at the operating-system level.
The \code{big.matrix} structure has several advantages such as
support of shared memory for efficiency in parallel computing, 
reference behavior that avoids unnecessary temporary copies of 
massive objects, and column-major format that is compatible with legacy 
linear algebra packages (e.g., \proglang{BLAS}, \proglang{LAPACK})
\citep{Kane:Emer:West:scal:2013}.
The package provides utility to read in a csv file to form 
a \code{big.matrix} object, but it only allows one type 
of data, numeric; it is a numeric matrix after all.

Package \pkg{ff} provides data structures that are stored in binary
flat files but behave (almost) as if they were in RAM by transparently 
mapping only a section (pagesize) of meta data in main memory.
Unlike \pkg{bigmemory}, it supports \proglang{R}'s standard atomic
data types (e.g., double or logical) as well as nonstandard, storage 
efficient atomic types (e.g., the 2-bit unsigned \code{quad} type allows
efficient storage of genomic data as a factor with levels A, T, G, and, C).
It also  provides class \code{ffdf} which is like \code{data.frame}
in \proglang{R}, and import/export filters for csv files.
A binary flat file can be shared by multiple \code{ff} objects in
the same or multiple \proglang{R} processes for parallel access.
Utility functions allow interactive process of selections of big data.

\subsubsection{Numerical Calculation}

The data management systems in packages \pkg{bigmemory} or \pkg{ff}
do not mean that one can apply existing \proglang{R} functions yet.
Even a simple statistical analysis such as linear model or survival 
analysis will need to be implemented for the new data structures.
Chunks of big data will be processed in RAM one at a time, and 
often, the process needs to be iterated over the whole data.
A special case is the linear model fitting, where one pass of the 
data is sufficient and no resetting from the beginning is needed.
Consider a regression model $E[Y] = X\beta$ with $n \times 1$ response 
$Y$, $n\times p$ model matrix $X$ and $p\times 1$ coefficient $\beta$.
The base \proglang{R} implementation \code{lm.fit} takes $O(np + p^2)$ 
memory, which can be reduced dramatically by processing in chunks.
The first option is to compute $X'X$ and $X'y$ in increment, and
get the least squares estimate of $\beta$, $\hat\beta = (X'X)^{-1}X'Y$.
This method is adopted in package \pkg{speedglm} \citep{Rpkg:speedglm}. 
A slower but more accurate option is to compute the incremental 
QR decomposition \citep{Mill:algo:1992} of $X = QR$ to get $R$ 
and $Q'Y$, and then solve $\beta$ from $R\beta = Q'Y$. 
This option is implemented in package \pkg{biglm} \citep{Rpkg:biglm}. 
Function \code{biglm} uses only $p^2$ memory of $p$ variables and
the fitted object can be updated with more data using \code{update}.
The package also provides an incremental computation of sandwich
variance estimator by accumulating a $(p + 1)^2\times (p + 1)^2$ 
matrix of products of $X$ and $Y$ without a second pass of the data.

In general, a numerical calculation needs an iterative algorithm
in computation and, hence, multiple passes of the data are necessary.
For example, a GLM fitting is often obtained through
the iterated reweighted least squares (IRLS) algorithm.
The \code{bigglm} function in package \pkg{biglm} implements the
generic IRLS algorithm that can be applied to any specific data 
management system such as DBMS, \pkg{bigmemory}, or \pkg{ff}, provided 
that a function \code{data(reset = FALSE)} supplies the next chunk of 
data or zero-row data if there is no more, and \code{data(reset = TRUE)} 
resets to the beginning of the data for the next iteration.
Specific implementation of the \code{data} function for object of 
class \code{big.matrix} and \code{ffdf} are provided in package 
\pkg{biganalytics} \citep{Rpkg:biganalytics}
and \pkg{ffbase} \citep{Rpkg:ffbase}, respectively.

For any statistical analysis on big data making use of the
data management system, one would need to implement the necessary
numerical calculations like what package \pkg{biglm} does for GLM.
The family of \pkg{bigmemory} provides a collection of functions
for \code{big.matrix} objects:
\pkg{biganalytics} for basic analytic and statistical functions,
\pkg{bigtabulate} for tabulation operations \citep{Rpkg:bigtabulate},
and \pkg{bigalgebra} for matrix operation with the BLAS and LAPACK
libraries \citep{Rpkg:bigalgebra}.
Some additional functions for \code{big.matrix} objects are available 
from other contributed packages, such as \pkg{bigpca} for principal 
component analysis and single-value decomposition \citep{Rpkg:bigpca}, 
and \pkg{bigrf} for random forest \citep{Rpkg:bigrf}.
For \code{ff} objects, package \pkg{ffbase} provides basic
statistical functions \citep{Rpkg:ffbase}. 
Additional functions for \code{ff} objects are provided in other 
packages, with examples including \pkg{biglars} for least angle regression 
and LASSO \citep{Rpkg:biglars} and \pkg{PopGenome} for population
genetic and genomic analysis \citep{Pfei:popg:2014}.

If some statistical analysis, such as generalized estimating equations
or Cox proportional hazards model, has not been implemented for big data, 
then one will need to modify the existing algorithm to implement it.
As pointed out by \citet[p.5]{Kane:Emer:West:scal:2013}, this would open
Pandora's box of recoding which is not a long-term solution for scalable
statistical analyses; this calls for redesign of the next-generation
statistical programming environment which could provide seamless
scalability through file-backed memory-mapping for big data, help
avoid the need for specialized tools for big data management, and allow
statisticians and developers to focus on new methods and algorithms.

\subsection{Breaking the Computing Power Barrier}
\label{sec:R:comp}

\subsubsection{Speeding Up}

As a high level interpreted language, for which most of instructions
are executed directly, \proglang{R} is infamously slow with loops.
Some loops can be avoided by taking advantage of the vectorized
functions in \proglang{R} or by clever vectorizing with some effort. 
When vectorization is not straightforward or loops are unavoidable, as in 
the case of MCMC,  acceleration is much desired, especially for big data.
The least expensive tool in a programmer's effort to speed up \proglang{R} 
code is to compile them to byte code with the \pkg{compiler} package,
which was developed by Luke Tierney and is now part of base \proglang{R}.
The byte code compiler translates the high-level \proglang{R} 
into a very simple language that can be interpreted by a 
very fast byte code interpreter, or virtual machine. 
Starting with \proglang{R}~2.14.0 in 2011, the base and recommended 
packages were pre-compiled into byte-code by default.
Users' functions, expressions, scripts, and packages can be compiled
for an immediate boost in speed by a factor of 2 to 5.

Computing bottlenecks can be implemented in a compiled language such as
\proglang{C/C++} or \proglang{FORTRAN} and interfaced to \proglang{R}
through \proglang{R}'s foreign language interfaces \citep[ch.5]{Rext}.
Typical bottlenecks are loops, recursions, and complex data structures.
Recent developments have made the interfacing with \proglang{C++} 
much easier than it used to be \citep{Edde:seam:2013}.
Package \pkg{inline} \citep{Rpkg:inline} provides functions that 
wrap \proglang{C/C++} (or \proglang{FORTRAN}) code as strings in
\proglang{R} and takes care of compiling, linking, and loading
by placing the resulting dynamically-loadable object code in
the per-session temporary directory used by \proglang{R}.
For more general usage, package \pkg{Rcpp} \citep{Edde:etal:rcpp:2011} 
provides \proglang{C++} classes for many basic \proglang{R} data 
types, which allow straightforward passing of data in both directions.
Package \pkg{RcpEigen} \citep{Rpkg:rcppeigen} provides access to the 
high-performance linear algebra library \proglang{Eigen} for a wide variety 
of matrix methods, various decompositions and support of sparse matrices. 
Package \pkg{RcppArmadillo} \citep{Edde:Sand:rcpp:2014} connects 
\proglang{R} with \proglang{Armadillo}, a powerful templated linear algebra 
library which provides a good balance between speed and ease of use.
Package \pkg{RInside} \citep{Rpkg:rinside} gives easy access of
\proglang{R} objects from \proglang{C++} by wrapping the existing
\proglang{R} embedding application programming interface (API) in
\proglang{C++} classes. 
The \pkg{Rcpp} project has revolutionized the integration of \proglang{R}
with \proglang{C++}; it is now used by hundreds of \proglang{R} packages.

Diagnostic tools can help identify the bottlenecks in \proglang{R} code.
Package \pkg{microbenchmark} \citep{Rpkg:microbenchmark} provides 
very precise timings for small pieces of source code, making it 
possible to compare operations that only take a tiny amount of time.
For a collection of code, run-time of each individual operation can
be measured with realistic inputs; the process is known as profiling.
Function \code{Rprof} in \proglang{R} does the profiling, but the
outputs are not intuitive to understand for many users.
Packages \pkg{proftools} \citep{Rpkg:proftools} and \pkg{aprof}
\citep{Rpkg:aprof} provide tools to analyze profiling outputs.
Packages \pkg{profr} \citep{Rpkg:profr}, 
\pkg{lineprof} \citep{Rpkg:lineprof}, and
\pkg{GUIProfiler} \citep{Rpkg:GUIProfiler}
provide visualization of profiling results.

\subsubsection{Scaling Up}

The \proglang{R} package system has long embraced integration of parallel 
computing of various technologies to address the big data challenges.
For embarrassingly parallelizable jobs such as bootstrap or simulation, 
where there is no dependency or communication between parallel tasks,
many options are available with computer clusters or multicores.
\citet{schm:etal:stat:2009} reviewed the then state-of-the-art parallel
computing with \proglang{R}, highlighting two packages for cluster use:
\pkg{Rmpi} \citep{Yu:rmpi:2002} which provides an \proglang{R} interface
to the Message Passing Interface (MPI) in parallel computing;
\pkg{snow} \citep{Ross:Tier:Li:simp:2007} which provides an abstract 
layer with the communication details hidden from the end users.
Since then, some packages have been developed and some discontinued.
Packages \pkg{snowFT} \citep{Rpkg:snowFT} and \pkg{snowfall} 
\citep{Rpkg:snowfall} extend \pkg{snow} with fault tolerance and
wrappers for easier development of parallel \proglang{R} programs.
Package \pkg{multicore} \citep{Rpkg:multicore} provides parallel 
processing of \proglang{R} code on machines with multiple cores or CPUs.
Its work and some of \pkg{snow} have been incorporated into the 
base \proglang{R} package \pkg{parallel},
which was first included in \proglang{R}~2.14.0 in 2011.
Package \pkg{foreach} \citep{Rpkg:foreach} allows general iteration 
over elements in a collection without any explicit loop counter.
Using \code{foreach} loop without side effects facilitates executing 
the loop in parallel with different parallel mechanisms, including
those provided by \pkg{parallel}, \pkg{Rmpi}, and \pkg{snow}.
For massive data that exceed the computer memory, a combination of
\pkg{foreach} and \pkg{bigmemory}, with shared-memory data 
structure referenced by multiple processes, provides a framework 
with ease of development and efficiency of execution (both in speed
and memory) as illustrated by \citet{Kane:Emer:West:scal:2013}.
Package \pkg{Rdsm} provides facilities for distributed shared memory
parallelism at the \proglang{R} level, and combined with \pkg{bigmemory},
it enables parallel processing on massive, out-of-core matrices.

The ``Programming with Big Data in \proglang{R}'' project (pbdR) enables 
high-level distributed data parallelism in \proglang{R} with easy 
utilization of large clusters with thousands of cores \citep{pbdR2012}.
Big data are interpreted quite literally to mean that a dataset requires
parallel processing either because it does not fit in the memory of a 
single machine or because its processing time needs to be made tolerable.
The project focuses on distributed memory systems where data are distributed 
across processors and communications between processors are based on MPI.
It consists of a collection of \proglang{R} packages in a hierarchy.
Package \pkg{pbdMPI} provides \code{S4} classes to directly interface
with MPI to support the Single Program Multiple Data (SPMD) parallelism.
Package \pkg{pbdSLAP} serves as a mechanism to utilize a subset of
functions of scalable dense linear algebra in \proglang{ScaLAPACK} 
\citep{Blac:etal:scal:1997},
a subset of \proglang{LAPACK} routines redesigned with the SPMD style.
Package \pkg{pbdBASE} contains a set of wrappers of low level functions
in \proglang{ScaLAPACK}, upon which package \pkg{pbdMAT} builds to
provide distributed dense matrix computing while preserving the friendly
and familiar \proglang{R} syntax for these computations.
Demonstrations on how to use these and other packages from the pbdR
are available in package \pkg{pbdDEMO}.

A recent, widely adopted open source framework for massive data 
storage and distributed computing is \proglang{Hadoop} \citep{Hadoop}.
Its heart is an implementation of the MapReduce programming model
first developed at Google \citep{Dean:Ghem:mapr:2008},
which divides the data to distributed systems and computes for each 
group (the map step), and then recombines the results (the reduce step).
It provides fault tolerant and scalable storage of massive
datasets across machines in a cluster \citep{Whit:hado:2011}.
The model suits perfectly the embarrassingly parallelizable jobs
and the distributed file system helps break the memory boundary.
\citet[ch.5--8]{McCa:West:para:2011} demonstrated three ways to 
combine \proglang{Hadoop} and \proglang{R}.
The first is to submit \proglang{R} scripts directly to a
\proglang{Hadoop} cluster, which gives the user the most control and
the most power, but comes at the cost of a \proglang{Hadoop} learning
curve. The second is a pure \proglang{R} solution via package
\pkg{Rhipe}, which hides the communications to \proglang{Hadoop} from
\proglang{R} users. The package (not on CRAN) is from the
\proglang{RHIPE} project, which stands for \proglang{R} and
\proglang{Hadoop} Integrated Programming Environment
\citep{Guha:etal:larg:2012}.
With \pkg{Rhipe}, data analysts only need to write \proglang{R} code
for the map step and the reduce step \citep{Guha:etal:larg:2012}, and
get the power of \proglang{Hadoop} without leaving \proglang{R}.
The third approach targets specifically the Elastic MapReduce (EMR) 
at Amazon by a CRAN package \pkg{segue} \citep{Rpkg:segue}, 
which makes EMR as easy to use as a parallel backend for 
\code{lapply}-style operations. 
An alternative open source project that connects \proglang{R} and
\proglang{Hadoop} is the RHadoop project, which is actively being
developed by Revolution Analytics \citep{RHadoop}.
This project is a collection of \proglang{R} packages that allow users
to manage and analyze data with \proglang{Hadoop}: 
\pkg{rhbase} provides functions for database management for the 
HBase distributed database, \pkg{rhdfs} provides functions for
\proglang{Hadoop} distributed file system (HDFS), \pkg{rmr} provides
functions to \proglang{Hadoop} MapReduce functionality, \pkg{plymr}
provides higher level data processing for structured data, and the
most recent addition \pkg{ravro} provides reading and writing
functions for files in \code{avro} format, an efficient data
serialization system developed at Apache \citep{Avro}.

\proglang{Spark} is a more recent, cousin project of \proglang{Hadoop}
that supports tools for big data related tasks \citep{Spark}.
The functions of \proglang{Spark} and \proglang{Hadoop} are neither
the exactly same nor mutually exclusive, and they often work together.
\proglang{Hadoop} has its own distributed storage system, which
is fundamental for any big data computing framework, allowing vast 
datasets to be stored across the hard drives of a scalable computer
cluster rather than on a huge costly hold-it-all device.
It persists back to the disk after a map or reduce action.
In contrast, \proglang{Spark} does not have its own distributed file
system, and it processes data in-memory \citep{Zaha:etal:2010}.
The biggest difference is disk-based computing versus memory-based
computing. This is why \proglang{Spark} could work 100 times faster
than hadoop for some applications when the data fit in the memory.
Some applications such as machine learning or stream processing where
data are repeatedly queried makes \proglang{Spark} an ideal framework.
For big data that does not fit in memory, \proglang{Spark}'s operators
spill data to disk, allowing it to run well on any sized data. For
this purpose, it can be installed on top \proglang{Hadoop} to take
advantage of \proglang{Hadoop}'s HDFS.
An \proglang{R} frontend to \proglang{Spark} is provided in
\proglang{R} package \pkg{SparkR} \citep{Rpkg:SparkR}, which has
become part of Apache \proglang{Spark} recently.
By using \proglang{Spark}'s distributed computation engine, the
package allows users to run large scale data analysis such as
selection, filtering, aggregation from \proglang{R}.
\citet{Kara:etal:2015} provides a summary of the state-of-the-art on
using \proglang{Spark}.

As multicores have become the standard setup for computers today, 
it is desirable to automatically make use of the cores in implicit 
parallelism without any explicit requests from the user.
The experimental packages \pkg{pnmath} and \pkg{pnmath0} by Luke Tierney
replace a number of internal vector operations in \proglang{R} with
alternatives that can take advantage of multicores \citep{Tier:code:2009}.
For a serial algorithm such as MCMC, it is desirable to 
parallelize the computation bottleneck if possible, but this generally
involves learning new computing tools and the debugging can be challenging.
For instance, \citet{Yan:etal:para:2007} used the parallel linear algebra
package (PLAPACK) \citep{Geji:usin:1997} for the matrix operations 
(especially the Cholesky decomposition) in a MCMC algorithm for Bayesian
spatiotemporal geostatistical models, but the scalability was only moderate.

When random numbers are involved as in the case of simulation, 
extra care is needed to make sure the parallelized jobs run 
independent (and preferably reproducible) random-number streams.
Package \pkg{rsprng} \citep{Rpkg:rsprng} provides an interface to the
Scalable Parallel Random Number Generators (SPRNG) \citep{Masc:Srin:algo:2000}.
Package \pkg{rlecuyer} \citep{Rpkg:rlecuyer} provides an interface to 
the random number generator with multiple independent streams developed 
by \citet{L'Ec:etal:2002}, the ideas of which are also implemented in
the base package \pkg{parallel}: make independent streams by separating 
a single stream with a sufficiently large number of steps apart.
Package \pkg{doRNG} \citep{Rpkg:doRNG} provides functions to perform
reproducible parallel \code{foreach} loops, independent of the 
parallel environment and associated \code{foreach} backend.

From a hardware perspective, many computers have mini clusters of
graphics processing units (GPUs) that can help with bottlenecks.
GPUs are dedicated numerical processors that were originally 
designed for rendering three dimensional computer graphics. 
A GPU has hundreds of processor cores on a single chip and can be 
programmed to apply the same numerical operations on large data array. 
\citet{Such:etal:unde:2010} investigated the use of GPUs in massively
parallel massive mixture modeling, and showed better performance
of GPUs than multicore CPUs, especially for larger samples.
To reap the advantage, however, one needs to learn the related 
tools such as Compute Unified Device Architecture (CUDA),
Open Computing Language (OpenCL), and so on, which may be prohibitive.
An \proglang{R} package \pkg{gputools} \citep{Rpkg:gputools} 
provides interface to NVidia CUDA toolkit and others.

If one is willing to step out of the comfort zone of \proglang{R} and
take full control/responsibility of parallel computing, one may
program with open source MPI or Open Multi-Processing (OpenMP).
MPI is a language-independent communication system designed for
programming on parallel computers, targeting high performance,
scalability and portability \citep{Pach:para:1997}.
Most MPI implementations are available as libraries from
\proglang{C/C++}, \proglang{FORTRAN},  and any language that
can interface with such libraries, including \proglang{C\#},
\proglang{Java} or \proglang{Python}. 
The interface from \proglang{R} can be accessed with package
\pkg{Rmpi} \citep{Yu:rmpi:2002} as mentioned earlier.
Freely available implementations include OpenMPI (not OpenMP)
and MPICH, while others come with license such as Intel MPI.
OpenMP is an API that supports multi-platform shared memory
multiprocessing programming in \proglang{C/C++} and 
\proglang{FORTRAN} on most processor architectures and 
operating systems \citep{chapman2008using}.
It is an add on to compilers (e.g., \proglang{gcc}, intel
compiler) to take advantage of of shared memory systems such as
multicore computers where processors shared the main memory.
MPI targets both distributed as well as shared momory systems while
OpenMP targets only shared memory systems.
MPI provides both process and thread based approach while OpenMP
provides only thread based parallilism.
OpenMP uses a portable, scalable model that gives programmers a simple
and flexible interface for writing multi-threaded programs in
\proglang{C/C++} and \proglang{FORTRAN} \citep{Dagu:Enon:open:1998}. 
Debugging parallel programs can be very challenging.

\section{Commercial Statistical Software}
\label{sec:comm}

\proglang{RRE} is the core product of Revolution Analytics 
(formerly Revolution Computing), a company 
that provides \proglang{R} tools, support, and training. 
\proglang{RRE} focuses on big data, large scale multiprocessor 
(or high performance) computing, and multicore functionality.
Massive datasets are handled via EMA and parallel EMA (PEMA)
when multiprocessors or multicores are available.
The commercial package \pkg{RevoScaleR} \citep{Rpkg:RevoScaleR} 
breaks the memory boundary by a special \code{XDF} data format 
that allows efficient storage and retrieval of data.
Functions in the package (e.g., \code{rxGlm} for GLM fitting) 
know to work on a massive dataset one chunk at a time.
The computing power boundary is also addressed --- functions 
in the package can exploit multicores or computer clusters.
Packages from the aforementioned open source project RHadoop 
developed by the company provide support for \proglang{Hadoop}.
Other components in \proglang{RRE} allow high speed connection for various 
types of data sources and threading and inter-process 
communication for parallel and distributed computing.
The same code works on small and big data, and on workstations, 
servers, clusters, \proglang{Hadoop}, or in the cloud.
The single workstation version of \proglang{RRE} is free for academic
use currently, and was used in the case study in Section~\ref{sec:case}.

\proglang{SAS}, one of the most widely used commercial software
for statistical analysis, provides big data support through SAS 
High Performance Analytics.
Massive datasets are approached by grid computing, in-database 
processing, in-memory analytics and connection to \proglang{Hadoop}.
The SAS High Performance Analytics Products include statistics, 
econometrics, optimization, forecasting, data mining, and text mining,
which, respectively, correspond to SAS products STAS, ETS, OR, 
high-performance forecasting, enterprise miner, and text miner
\citep{Cohe:Rodr:high:2013}.

IBM \proglang{SPSS}, the Statistical Product and Services Solution,
provides big data analytics through SPSS Modeler, SPSS Analytic 
Server, SPSS Collaboration and Deployment Services, and SPSS 
Analytic Catalyst \citep{SPSS}.
SPSS Analytic Server is the foundation and it focuses on high performance
analytics for data stored in Hadoop-based distributed systems. 
SPSS modeler is the high-performance data mining workbench, utilizing
SPSS Analytic Server to leverage big data in Hadoop environments. 
Analysts can define analysis in a familiar and accessible workbench to 
conduct analysis modeling and scoring over high volumes of varied data. 
SPSS Collaboration and Deployment Services helps manage analytical assets, 
automate processes and efficiently share results widely and securely.
SPSS Analytic Catalyst is the automation of analysis that makes
analytics and data more accessible to users.

\proglang{MATLAB} provides a number of tools to tackle the
challenges of big data analytics \citep{Matlab}.
Memory mapped variables map a file  or a proportion of a file to a 
variable in RAM; disk variables direct access to variables from files 
on disk; datastore allows access to data that do not fit into RAM.
Their combination addresses the memory boundary.
The computation power boundary is broken by intrinsic multicore math, GPU 
computing, parallel computing, cloud computing, and \proglang{Hadoop} support.

\section{A Case Study}
\label{sec:case}

The airline on-time performance data from the 2009 ASA Data Expo 
(\url{http://stat-computing.org/dataexpo/2009/the-data.html})
is used as a case study to demonstrate a logistic model fitting 
with a massive dataset that exceeds the RAM of a single computer.
The data is publicly available and has been used for demonstration
with big data by \citet{Kane:Emer:West:scal:2013} and others.
It consists of flight arrival and departure details for all 
commercial flights within the USA, from October 1987 to April 2008. 
About 12 million flights were recorded with 29 variables.
A compressed version of the pre-processed data set from the bigmemory
project (\url{http://data.jstatsoft.org/v55/i14/Airline.tar.bz2})
is approximately 1.7GB, and it takes 12GB when uncompressed.

The response of the logistic regression is late arrival which was 
set to 1 if a flight was late by more than 15 minutes and 0 otherwise.
Two binary covariates were created from the departure time:
night (1 if departure occurred between 8pm and 5am) and
weekend (1 if departure occurred on weekends and 0 otherwise).
Two continuous covariates were included: departure hour (DepHour, 
range 0 to 24) and distance from origin to destination (in 1000 miles).
In the raw data, the departure time was an integer of the HHmm format.
It was converted to minutes first to prepare for DepHour.
Three methods are considered in the case study: 
1) combination of \code{bigglm} with package \pkg{bigmemory};
2) combination of \code{bigglm} with package \pkg{ff}; and
3) the academic, single workstation version of \proglang{RRE}.
The default settings of \pkg{ff} were used. 
Before fitting the logistic regression, the 12GB raw data needs to
be read in from the csv format, and new variables needs to be generated.
This leads to a total of $120,748,239$ observations with no missing data.
The \proglang{R} scripts for the three methods are in the supplementary
materials for interested readers.

\begin{table}[tbp]
\centering
\caption{Timing results (in seconds) for reading in the whole 12GB data, transforming to create new variables, and fitting the logistic regression with three methods: \pkg{bigmemory}, \pkg{ff}, and \proglang{RRE}.}
\label{tab:timing}
\begin{tabular}{rrrr}
  \toprule
 & Reading & Transforming & Fitting \\
  \midrule
\pkg{bigmemory}  &  968.6 & 105.5 & 1501.7  \\
\pkg{ff}         & 1111.3 & 528.4 & 1988.0 \\
\proglang{RRE}   &  851.7 & 107.5 &  189.4 \\
\bottomrule
\end{tabular}
\end{table}

The \proglang{R} scripts were executed in batch mode on a 8-core machine 
running CenOS (a free Linux distribution functionally compatible with 
Red Hat Enterprise Linux which is officially supported by \proglang{RRE}),
with Intel Core i7 2.93GHz CPU, and 16GB memory.
Table~\ref{tab:timing} summarizes the timing results of reading in the
whole 12GB data, transforming to create new variables, and fitting the 
logistic regression with the three methods.
The chunk sizes were set to be 500,000 observations for all three methods.
For \proglang{RRE}, this was set when reading in the data to the 
\code{XDF} format; for the other two methods, this was set at 
the fitting stage using function \code{bigglm}.
Under the current settings, \proglang{RRE} has a clear advantage in 
fitting with only 8\% of the time used by the other two approaches.
This is a result of the joint force of its using all 8~cores
implicitly and efficient storage and retrieval of the data;
the \code{XDF} version of the data is about 1/10 of the size 
of the external files saved by \pkg{bigmemory} or \pkg{ff}.
Using \pkg{bigmemory} and using \pkg{ff} in \code{bigglm} had 
very similar performance in fitting the logistic regression, but
the former took less time in reading, and significantly less
time (only about 1/5) in transforming variables of the latter.
The \pkg{bigmemory} method was quite close to the \proglang{RRE} 
method in the reading and the transforming tasks.
The \pkg{ff} method took longer in reading and transforming than the 
\pkg{bigmemory} method, possibly because it used much less memory. 

\begin{table}[tbp]
\centering
\caption{Logistic regression results for late arrival.}
\label{tab:fit}
\begin{tabular}{crc}
  \toprule
            & Estimate   & Std. Error ($\times 10^4$) \\
  \midrule
(Intercept) &   $-$2.985 &   9.470\\
DepHour     &      0.104 &   0.601\\
Distance    &      0.235 &   4.032\\
Night       &   $-$0.448 &   8.173\\
Weekend     &   $-$0.177 &   5.412\\
\bottomrule
\end{tabular}
\end{table}

The results of the logistic regression are identical from all
methods, and are summarized in Table~\ref{tab:fit}.
Flights with later departure hour or longer distance are
more likely to be delayed.
Night flights or weekend flights are less likely to be delayed.
Given the huge sample size, all coefficients were highly significant.
It is possible, however, that p-values can still be useful.
A binary covariate with very low rate of event may still
have an estimated coefficient with a not-so-low p-value
\citep{Schifano2015}, an effect only estimable with big data.

\begin{table}[tbp]
\centering
\caption{Time results (in seconds) for parallel computing quantiles of departure delay for each day of the week with 1 to 8 cores using \pkg{foreach}.}
\label{tab:para}
\begin{tabular}{rrrrrrrrr}
  \toprule
  & 1 &  2 &  3 &  4 &  5 &  6 &  7 &  8  \\
  \midrule
  \pkg{bigmemory} & 22.1 & 11.2 & 7.8 & 6.9 &  6.2 & 6.3 &  6.4 &  6.8\\
  \pkg{ff} &   21.4 & 11.0 & 7.1 &  6.7 & 5.8 & 5.9 & 6.1 & 6.8 \\
  \bottomrule
\end{tabular}
\end{table}

As an illustration of \pkg{foreach} for embarrassingly parallel
computing, the example in \citet{Kane:Emer:West:scal:2013} is 
expanded to include both \pkg{bigmemory} and \pkg{ff}.
The task is to find three quantiles (0.5, 0.9, and 0.99)
of departure delays for each day of the week;
that is, 7 independent jobs can run on 7 cores separately.
To make the task bigger, each job was set to run twice.
The resulting 14 jobs were parallelized with \code{foreach} on
the same Linux machine using 1 to 8 cores for the sake of illustration.
The \proglang{R} script is included in the supplementary materials.
The timing results are summarized in Table~\ref{tab:para}. 
There is little difference between the two implementations.
When there is no communication overhead, with 14 jobs one would
expect the run time to reduce to 1/2, 5/14, 4/14, 3/14, 3/14, 2/14,
and 2/14, respectively, with 2, 3, 4, 5, 6, 7 and 8~cores.
The impact of communication cost is obvious in Table~\ref{tab:para}.
The time reduction is only closer to the expectation in the ideal
case when the number of cores is smaller.

\section{Discussion}
\label{sec:disc}

This article presents a recent snapshot on statistical analysis with big data
that exceed the memory and computing capacity of a single computer.
Albeit under-appreciated by the general public or even mainstream academic
community, computational statisticians have made respectable progress
in extending standard statistical analysis to big data, with the most
notable achievements in the open source \proglang{R} community.
Packages \pkg{bigmemory} and \pkg{ff} make it possible in principle
to implement any statistical analysis with their data structure.
Nonetheless, for anything that has not been already implemented 
(e.g., survival analysis, generalized estimating equations, 
mixed effects model, etc.), one 
would need to implement an EMA version of the computation task, which
may not be straightforward and may involve some steep learning curves.
\proglang{Hadoop} allows easy extension of algorithms that do not require 
multiple passes of the data, but such analyses are mostly descriptive.
An example is visualization, an important tool in exploratory analysis.
With big data, the bottleneck is the number of pixels in the screen.
The bin-summarize-smooth framework for visualization of large data of
\citet{Wick:bin:2014} with package \pkg{bigvis} \citep{Rpkg:bigvis} 
may be adapted to work with \proglang{Hadoop}.

Big data present challenges much further beyond the territory of 
classic statistics, requiring joint workforce with domain knowledge, 
computing skills, and statistical thinking \citep{Yu:let:2014}.
Statisticians have much to contribute to both the intellectual vitality
and the practical utility of big data, but will have to expand their
comfort zone to engage high-impact, real world problems which are
often less structured or with ambiguity \citep{Jord:Lin:stat:2014}.
Examples are to provide structure for poorly defined problems, or
to develop methods/models for new types of data such as image or network.
As suggested by \citet{Yu:let:2014}, to play a critical role in the
arena of big data or own data science, statisticians need to work on 
real problems and relevant methodology and theory will follow naturally.

\section*{Acknowledgement}
The authors thank Stephen Archut, Fang Chen, and Joseph Rickert
for the big data analytics information on \proglang{SPSS}, 
\proglang{SAS}, and \proglang{RRE}.
An earlier version of the manuscript was presented at the 
``Statistical and Computational Theory and Methodology for Big Data 
Analysis'' workshop in February, 2014, at the 
Banff International Research Station in Banff, AB, Canada. 
The discussions and comments from the workshop participants are
gratefully acknowledged.

\section*{Supplementary Materials}
Four \proglang{R} scripts (and their outputs), along with
a descriptive README file are provided for the case study.
The first three are the logistic regression with, respectively,
combination of \pkg{bigmemory} with \code{bigglm} (\code{bigmemory.R}),
combination of \pkg{ff} with \code{bigglm} (\code{ff.R}), and
\proglang{RRE} (\code{RevR.R});
 their output files have \code{.Rout} extensions.
The first two run with \proglang{R}, while the third one needs \proglang{RRE}.
The fourth script is for the parallel computing with \pkg{foreach}
combined with \pkg{bigmemory} and \pkg{ff}, respectively.

\bibliographystyle{asa}
\bibliography{softrevTechRep}

\end{document}